\documentclass[sigconf]{acmart}

\usepackage{enumitem}
\usepackage{color,soul}
\usepackage{xcolor}
\usepackage{pifont}
\usepackage{xspace}
\usepackage{hyperref}
\usepackage{ulem}
\usepackage{multirow}
\usepackage{tabularx}
\usepackage{bm}
\usepackage{hyperref}

\usepackage{tikz}
\usetikzlibrary{calc}
\usepackage{subfigure}

\usepackage[table]{xcolor}

\usepackage{dcolumn}

\usepackage{algorithm}
\usepackage{algorithmic}

\usepackage{tikz}
\usepackage{subcaption} 

\AtBeginDocument{%
  }

\acmConference[EASE 2026]{The 30th International Conference on Evaluation and Assessment in Software Engineering}{9–12 June, 2026}{Glasgow, Scotland, United Kingdom}

\author{Jeremy Diamond}
\affiliation{%
  \institution{Universität Zürich}
  \city{Zurich}
  \country{Switzerland}
}
\email{diamond@ifi.uzh.ch}

\author{Vincenzo Stoico}
\affiliation{%
  \institution{Vrije Universiteit Amsterdam }
  \city{Amsterdam}
  \country{Netherlands}
}
\email{v.stoico@vu.nl}

\newcolumntype{d}[1]{D{.}{.}{#1}}  

\newcommand{\refkernel}{\texttt{Ref\_Rapl\allowbreak\_Kernel}\xspace} 
\newcommand{\refuser}{\texttt{Ref\_Rapl\allowbreak\_User}\xspace} 
\newcommand{\refbench}{\texttt{Ref\_Rapl\_Micro\_Bench}\xspace}

\newcommand{\RQone}{
  \noindent\textbf{$\mathbf{RQ_1}$} -- \textbf{What is the overhead of measuring software energy usage at 1 kHz using RAPL‑based tools?}
}

\newcommand{\RQtwo}{
    \noindent \textbf{$\mathbf{RQ_2}$} -- \textbf{To what extent do RAPL access mechanisms introduce overhead?}
}

\definecolor{open}{RGB}{236,0,255}
\definecolor{close}{RGB}{245,127,255}
\definecolor{read}{RGB}{0,255,215}
\definecolor{write}{RGB}{0,137,255}
\definecolor{sleep}{RGB}{1,158,80}
\definecolor{native}{RGB}{139,255,0}
\definecolor{customblue}{RGB}{69,183,209}   
\definecolor{brightBackground}{RGB}{236,248,250} 

\newcommand{\marktext}[2]{\colorbox{#1}{\strut #2}}

\definecolor{large}{rgb}{1, 0.7, 0.7} 
\definecolor{medium}{rgb}{1, 0.6, 0.3} 
\definecolor{small}{rgb}{1, 0.9, 0} 

\begin{document}
\title{What Is the Cost of Energy Monitoring? An Empirical Study on the Overhead of RAPL-Based Tools}
\begin{abstract}
The Running Average Power Limit (RAPL) interface is widely used to estimate software energy consumption via CPU and DRAM counters, but tool design differences and high-frequency polling can introduce measurement overhead, namely, extra time and energy consumed by the tool itself.
This paper quantifies the impact of RAPL-based tools on high-frequency (1\,kHz) energy monitoring and investigates mitigation strategies. We conduct two controlled experiments: the first evaluates seven tools, including a user-space application and a kernel module developed by the authors, against a no-tool baseline, using six NAS Benchmark functions to quantify overhead. The second experiment isolates and times key functions for polling Model-Specific Registers (MSRs) (\texttt{rdmsr} and \texttt{sys/proc\_read}) to estimate their execution latencies and identify potential slowdowns.
The results show that existing user-space tools can introduce substantial time overhead at 1 kHz, whereas our tools significantly reduce system call overhead and inline math overhead. The time overhead of existing tools ranges from 0.25\% to 46.75\%. Our solutions maintain time overhead levels close to the baseline. We also find that system calls are slower than \texttt{rdmsr}, which in turn is slower than traditionally long-running instructions like \texttt{cpuid}. These findings indicate that RAPL-based energy measurement can be substantially improved by simplifying tool design and employing lower-level instructions to access RAPL values.
Our findings provide guidance for practitioners on how to develop high-frequency energy profiling tools, show possible situations that can skew energy values, and demonstrate that access to RAPL values can be faster using specific techniques.
\end{abstract}


\maketitle

\section{Introduction}
Since its introduction in 2011 with Intel's Sandy Bridge microarchitecture, the Running Average Power Limit (RAPL) interface has become the de facto mechanism for estimating energy consumption in modern processors.
RAPL provides power consumption estimates for various hardware domains, including the CPU packages, cores, and DRAM, by leveraging on-chip monitoring sensors \cite{khan2018rapl}.
These estimates are made available to software through a set of Model-Specific Registers (MSRs), which can be accessed via the operating system kernel.
The kernel exposes these MSRs through dedicated files in the power management interface of the system (e.g., through the powercap framework), enabling user-level tools to read and analyze energy usage data efficiently \cite{zhang2021red, phung2018modeling, venkatesh2013evaluation}.
The introduction of RAPL led to a surge in the number of tools leveraging it to allow users to query energy consumption through the RAPL counters, such as Scaphandre \cite{scaphandre}, PowerJoular \cite{noureddine-ie-2022}, CodeCarbon \cite{codecarbon}, and perf \cite{perf}.
Thus, \textit{RAPL became the key framework for empirically assessing software energy efficiency}.


RAPL-based tools offer distinct features and measurement scopes.
For example, Scaphandre focuses on monitoring the energy consumption of containers, whereas CodeCarbon measures energy usage at the code level and estimates the corresponding carbon footprint. Although both tools aim to profile software energy consumption, they differ in their design and implementation.  
Additionally, frequent polling required for sampling energy values can introduce measurement overhead, especially during fine-grained profiling, because exposing RAPL values from the kernel to user space involves inherently slow operations, such as system calls \cite{tanenbaum_modern_os}.
This overhead is typically negligible in common scenarios where the observed software runs for several seconds or longer, for example, when polling RAPL at a 1 Hz rate.
However, it may become significant and potentially distort measurements in experiments requiring high-frequency sampling, such as function-level energy analysis.
While measurement overhead is recognized in the literature \cite{van2025s, thamm2025strategies}, to the best of our knowledge, it has been explicitly quantified only by \citet{raffin2024dissecting}. We believe that the implementation complexity of energy profilers, along with the overhead introduced by the user-kernel function call chain, can contribute to measurement overhead, especially when sampling is performed at high frequency.

The \textbf{goal} of this study is to investigate how, and under which conditions, the use of RAPL-based tools for measuring software energy consumption may introduce measurement overhead.
Through two controlled experiments, we examine how both the implementation of existing tools and the process of retrieving energy data via RAPL affect the measurements.
The first experiment evaluates the overhead introduced by the implementation complexity of a tool.
This experiment evaluates a total of seven tools: five well-known solutions, Perf, PowerJoular, Turbostat, Scaphandre, and CodeCarbon, along with one user-space tool developed by the authors (\refuser) and another that leverages a custom kernel module also created by the authors (\refkernel).
The experiment is conducted with all tools configured to use a polling frequency of 1 kHz.
Since this frequency is not supported by all tools, namely PowerJoular, Scaphandre, and CodeCarbon, these tools were patched to enable it for the experiment.
All the tools profile six functions taken from the NAS Parallel Benchmarks \cite{araujo2023parallel}, which are a set of benchmarks developed by NASA to evaluate the performance of highly parallel supercomputers.
The second experiment analyzes the execution time of some representative operations involved in the retrieval of RAPL values from kernel to user space.
We implement a custom kernel module, called \refbench, to benchmark the execution time of RAPL-related \texttt{rdmsr} instructions and of the \texttt{proc\_read} and \texttt{sys\_read} system calls. These operations are compared to the execution time of a known long running one, \texttt{cpuid}, as well as the simple ones, \texttt{mov}, and \texttt{nop}, to provide a baseline for understanding the relative overhead introduced by more complex operations.
%

Across all six NAS benchmarks, the results show that our in-house tools, \refuser and \refkernel, maintain execution times very close to the no-tool baseline (No\_Tools), with \refuser often achieving the lowest relative overhead.
In contrast, commercial profilers such as Scaphandre and CodeCarbon can cause slowdowns when profiling at 1 kHz, reaching relative overheads up to +46.75\%. 
Absolute overheads confirm this pattern: deviations from the baseline for \refkernel and \refuser remain within a few seconds, while Scaphandre and CodeCarbon introduce much larger slowdowns.
Therefore, we suggest opting for tools with simple implementation and that can potentially reduce the length of the function call chain to retrieve RAPL energy values.
The results reveal clear performance differences between the \texttt{rdmsr} instructions and system calls.
\texttt{rdmsr} instructions show latencies in the range of approximately 2.3$\times$10$^{-4}$\,ms to 5.6$\times$10$^{-4}$\,ms, on the tested hardware, depending on the specific register accessed.
In contrast, system calls such as \texttt{sys\_read} require about 1.36$\times$10$^{-3}$\,ms per execution on that same hardware, roughly an order of magnitude slower than \texttt{rdmsr}.
This greater latency can arise from the transition between user and kernel space, which can slow down the exposure of RAPL energy values.


The main contributions are: an empirical study of RAPL-based tool overhead at 1 kHz; two alternative approaches (\refkernel and \refuser); a discussion of implications; and a replication package with all measurements, analysis scripts, and kernel module source code \cite{ReplicationPackage}. This work helps developers reduce measurement overhead and lays the groundwork for kernel module-based, high-frequency monitoring of software energy and quality attributes.

\section{Background}
\label{sec:background}
\subsection{RAPL}
RAPL (Running Average Power Limit) is an Intel hardware feature introduced in Sandy Bridge processors for monitoring and controlling energy consumption of Intel CPUs \cite{khan2018rapl}. It provides counters that track accumulated energy in microjoules, updated roughly every millisecond. RAPL divides the processor into power domains, each with dedicated energy counters accessible via model-specific registers (MSRs). Tools access RAPL counters either by reading Model-Specific Registers (MSRs) directly using the \texttt{rdmsr} instruction at addresses assigned to a specific domain, or by reading kernel-exposed files that provide RAPL energy values, such as the \texttt{/sys/class/powercap/intel-rapl} interface available since Linux kernel 3.13, which contains files for each RAPL domain. The covered domains include \textit{Package (PKG)}, which covers the entire CPU package including cores and uncore components; \textit{Core (PP0)}, which tracks energy used by CPU cores; \textit{Uncore (PP1)}, which typically accounts for the integrated GPU or nearby components; and \textit{DRAM}, representing the energy consumed by attached memory.

\subsection{RAPL-based Energy Profilers}
\begin{algorithm}
\footnotesize
\caption{RAPL Sampling Loop}\label{alg:rapl_loop}
\begin{algorithmic}[1]
    \WHILE{true}
        \STATE \textbf{Open output file}
        \STATE \textbf{Initialize RAPL interface}
        \STATE \textit{Record timestamp 1}
        \STATE \textit{Read RAPL registers (up to four syscalls)}
        \STATE \textit{Record timestamp 2}
        \STATE \textit{Compute energy delta}
        \STATE \textbf{Write results to output file}
        \STATE \textbf{Close RAPL interface}
        \STATE \textbf{Close output file}
        \STATE \textbf{Sleep 1 ms}
    \ENDWHILE
\end{algorithmic}
\end{algorithm}

All third-party tools examined in this comparison, excluding those developed by the authors, share a similar algorithmic structure \footnote{RAPL Implementation File - PowerJoular:  \href{https://github.com/joular/powerjoular/blob/develop/src/intel_rapl_sysfs.adb}{intel\_rapl\_sysfs.adb} and CodeCarbon: \href{https://github.com/mlco2/codecarbon/blob/master/codecarbon/core/rapl.py}{rapl.py}}
As explained in Section \ref{sec:subject}, our selection criteria focus on tools implementing sampling approaches that closely resemble \textit{Algorithm \ref{alg:rapl_loop}}, which outlines a typical RAPL-based energy measurement loop.
This loop design introduces two \textit{primary sources of overhead} that can skew measurement results. First, each iteration may open, write to, and close files, resulting in up to four extra system calls per measurement; although not all tools perform every redundant operation, each incurs at least one unnecessary syscall per iteration. Second, executing mathematical computations in real time (or in a parallel thread) increases CPU activity during sampling, thereby affecting both power and timing measurements \cite{shahid2021improving}. This effect can be minimized by separating sampling from calculation processing data at discrete intervals rather than during the active measurement loop.

\subsubsection{\refuser}

\begin{algorithm}
\footnotesize
\caption{\refuser\ Sampling Procedure}\label{alg:ref_rapl_U}
\begin{algorithmic}[1]
    \STATE \textbf{Open output file}
    \STATE \textbf{Initialize RAPL interface}
    \STATE \textit{int index = 0}
    \STATE \textit{sample\_type sample\_cache = 0}
    \WHILE{test is running}
        \STATE \textit{Record timestamp 1}
        \STATE \textit{Read RAPL registers (up to four syscalls)}
        \STATE \textit{Record timestamp 2}
        \STATE \textit{sample\_cache[index] = sample(ts1, ts2, rapl)}
        \IF{index == 100}
            \STATE \textbf{Write cache contents to output file}
            \STATE \textit{index = 0}
        \ENDIF
        \STATE \textbf{Sleep 1 ms}
    \ENDWHILE
    \STATE \textbf{Close RAPL interface}
    \STATE \textbf{Close output file}
\end{algorithmic}
\end{algorithm}

\refuser\ is our user-space energy profiling tool specifically designed to minimize the number of system calls performed during measurement.
Unlike conventional profilers, it opens and closes output files only once and uses an in-memory cache of samples allocated on the stack during initialization. Instead of writing each measurement immediately, \refuser accumulates samples in this cache and writes them collectively after every \(n\) samples, as shown in Algorithm \ref{alg:ref_rapl_U}. The current implementation uses a cache size of 128, though this value can be adjusted as needed. Because memory access is significantly faster than performing file I/O operations, this design dramatically reduces syscall overhead, approximately 6-9 system calls per iteration down to a theoretical 5.001 per loop.
Once data collection is complete, a separate offline process reads the recorded log and performs all necessary computations.
This deferred calculation strategy ensures that processing does not interfere with the measurements. 

\subsubsection{\refkernel}

\begin{algorithm}
\footnotesize
\caption{Ref\_Rapl\_Kernel Sampling Procedure}\label{alg:ref_rapl_K}
\begin{algorithmic}[1]
    \STATE \textbf{Open output file}
    \STATE \textbf{Initialize RAPL interface}
    \WHILE{test is running}
        \STATE \textbf{Read ring buffer via \texttt{/proc} file}
        \STATE \textbf{Append data to output file}
        \STATE \textbf{Sleep 0.1 s}
    \ENDWHILE
    \STATE \textbf{Close RAPL interface}
    \STATE \textbf{Close output file}
\end{algorithmic}
\end{algorithm}

\texttt{Ref\_Rapl\_Kernel} mirrors the design of \refuser but operates within the kernel.
In this approach, a cache of 128 samples is maintained in kernel space using a ring buffer.
While the buffer size can be adjusted, choosing a power of two allows for efficient, branchless index wrapping within the buffer.
The corresponding user-space program periodically reads this cached data through the \texttt{/proc} interface, typically every 100 milliseconds, and writes the complete buffer to an output file.
This process is illustrated in Algorithm \ref{alg:ref_rapl_K}.
The design dramatically reduces measurement overhead to only two system calls per 100 samples. 
By shifting the caching process into kernel space and reducing interaction frequency between user and kernel contexts, this design can achieve near-minimal syscall overhead.
The architectural choice to employ batched reads and optimized MSR access leads to cleaner sampling loops and improved profiling precision at negligible computational cost.

\subsubsection{\refbench}

\refbench is a kernel module designed to benchmark individual operations, including \texttt{rdmsr} instructions and system calls. It leverages the \texttt{proc} and \texttt{sys} virtual file systems to execute instruction loops of size \(n\) and measure their latency. Both virtual file systems create interfaces that, when read from or written to, trigger kernel-level functions.

In this setup, the directory \texttt{/proc/microbench} contains files that run loops of various instructions, including \texttt{nop}, \texttt{mov}, and \texttt{cpuid}, as well as RAPL-related \texttt{rdmsr}.
%
%
Additionally, two syscall benchmark files are provided: one under \texttt{/proc/microbench} and another in \texttt{/sys/kernel/syscall\_bench/}. These return pairs of timestamps recorded immediately before and after executing an \texttt{rdmsr}. By sampling and comparing these timestamps, users can subtract the measured \texttt{rdmsr} overhead and assess how system call latency compares to simple instructions, such as \texttt{nop}. 

\subsection{Tools Validation}
\label{sec:flame}
\begin{figure}
\centering
\subfigure[PowerJoular]{ 
    \label{fig:flamePower} 
    \includegraphics[width=\linewidth]{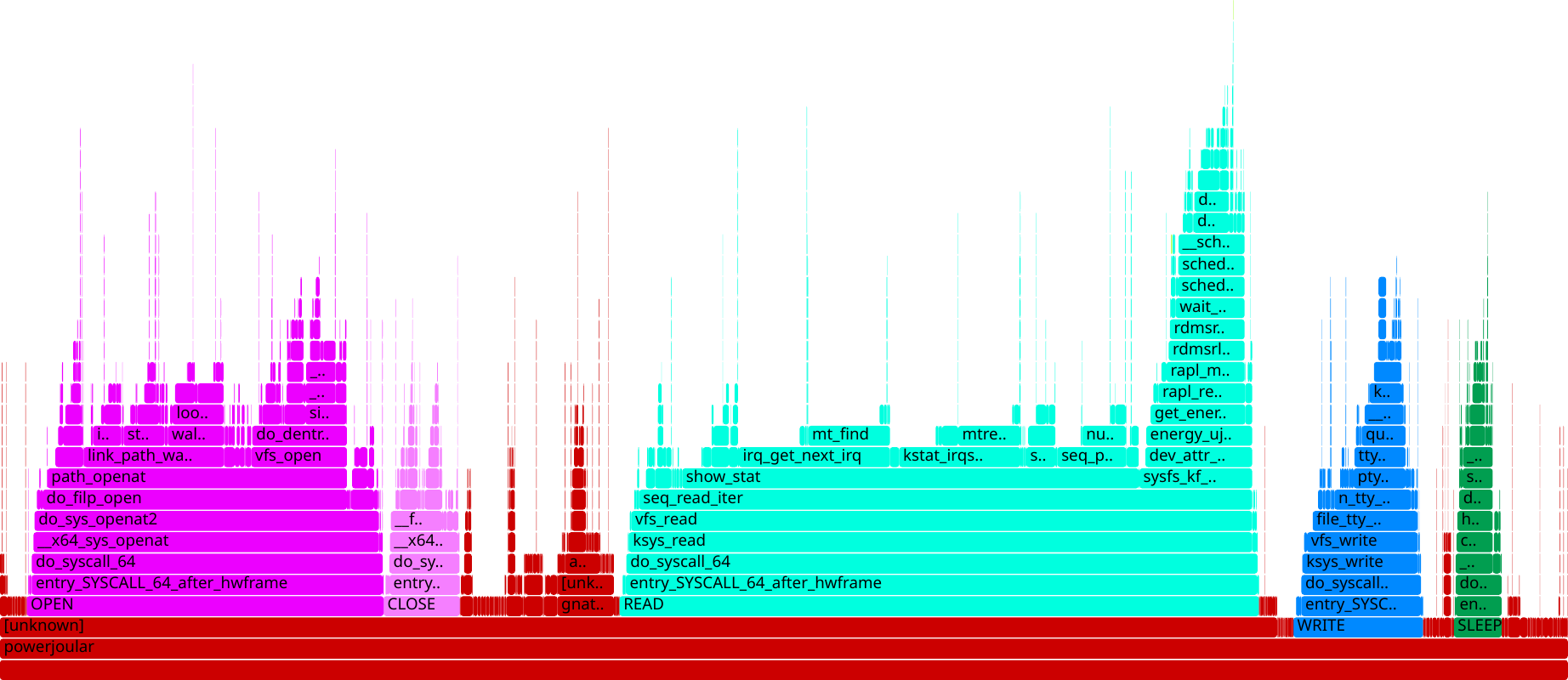}
    } 
\subfigure[\refuser]{ 
    \label{fig:flameUser} 
    \includegraphics[width=\linewidth]{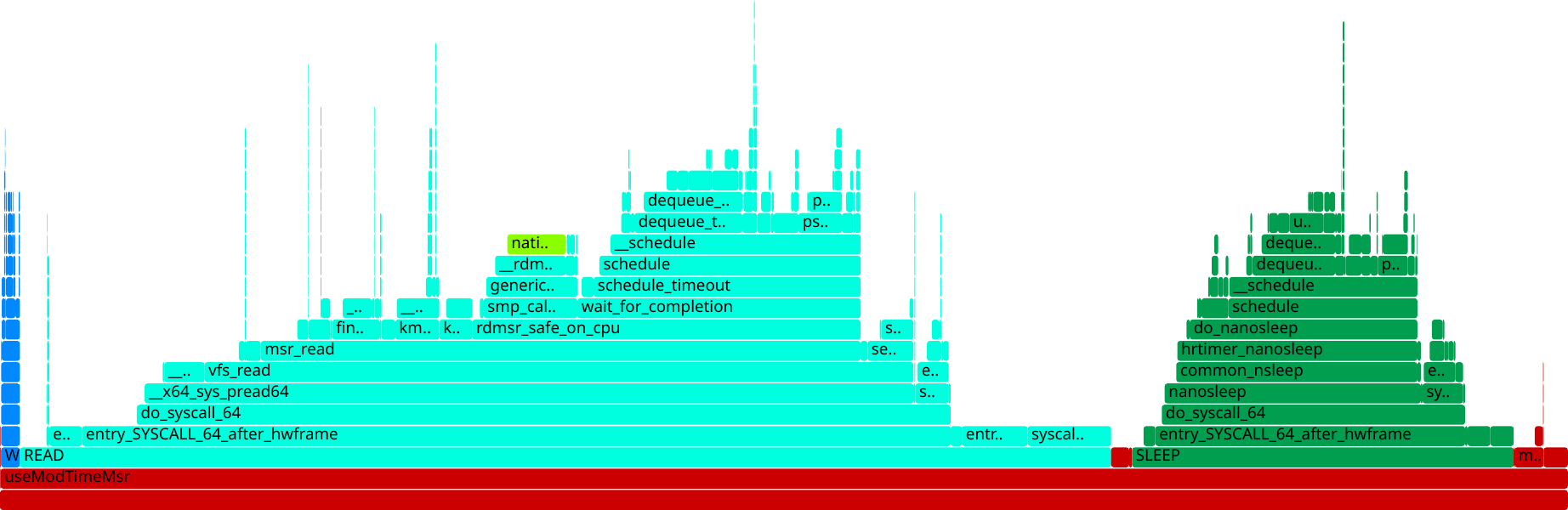} } 
\subfigure[\refkernel]{ 
    \label{fig:flameKernel} 
    \includegraphics[width=\linewidth]{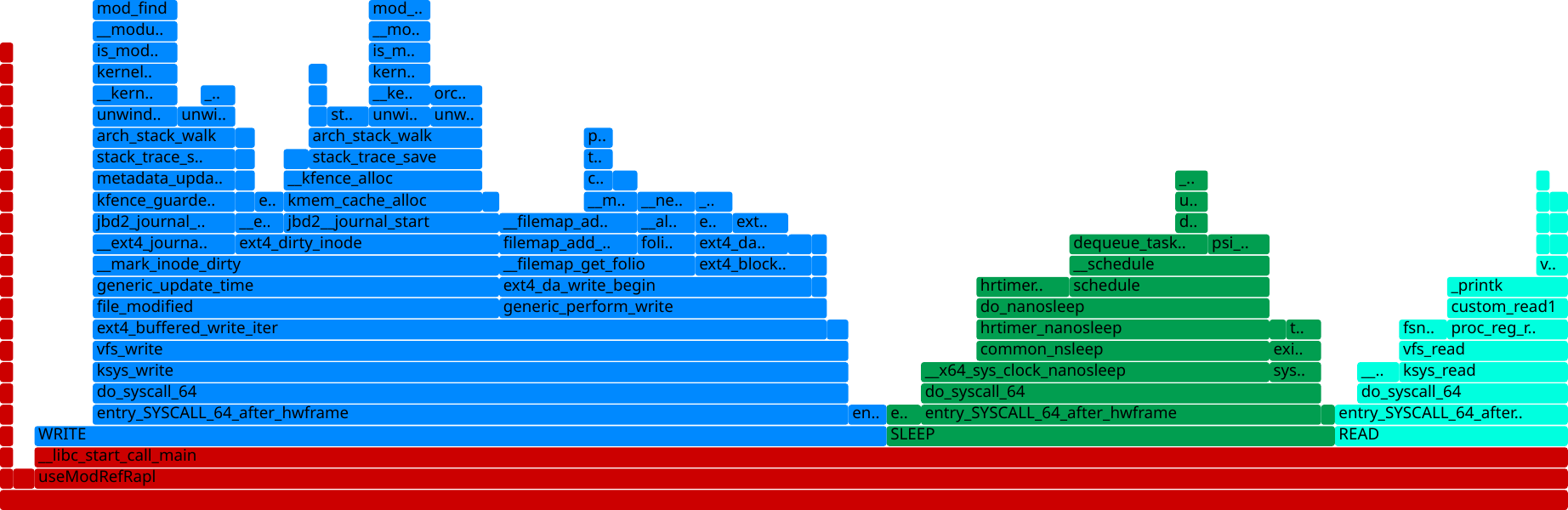} }
%



\caption{
    Flamegraphs comparison in percent of compute time spent on each task.
    \textbf{Color legend:} 
    \protect\tikz\protect\draw[fill={rgb,255:red,236;green,0;blue,255}] (0,0) rectangle (0.5em,0.5em); open,
    \protect\tikz\protect\draw[fill={rgb,255:red,245;green,127;blue,255}] (0,0) rectangle (0.5em,0.5em); close,
    \protect\tikz\protect\draw[fill={rgb,255:red,0;green,255;blue,215}] (0,0) rectangle (0.5em,0.5em); read,
    \protect\tikz\protect\draw[fill={rgb,255:red,0;green,137;blue,255}] (0,0) rectangle (0.5em,0.5em); write,
    \protect\tikz\protect\draw[fill={rgb,255:red,1;green,158;blue,80}] (0,0) rectangle (0.5em,0.5em); sleep,
    \protect\tikz\protect\draw[fill={rgb,255:red,139;green,255;blue,0}] (0,0) rectangle (0.5em,0.5em); rdmsr
}
\label{fig:flamegraphs}
\vspace{-.5cm}
\end{figure}

To validate the designs of \refuser and \refkernel, and to demonstrate how retrieving RAPL values can introduce overhead, we executed a 30-second sleep for each component as well as for PowerJoular. During these runs, we captured the user–kernel function call chains and measured the execution time of each function within the chain.
We visualize the resulting call chains using flame graphs \cite{FlameGraph}. Flame graphs are interactive SVG visualizations that illustrate how runtime is distributed across functions. The X-axis represents the proportion of total execution time spent in each function, while functions are stacked vertically on top of their callers. For instance, if \texttt{syscall\_read} invokes \texttt{native\_rdmsr}, the latter will appear above the former in the flame graph.




PowerJoular can be used to generate the flame graph because it is a compiled native binary (in Ada using GCC), making it usable with system-level profilers like \texttt{perf}.
Figure~\ref{fig:flamePower} highlights all major function calls of interest. It shows that 22.76\% of execution cycles are spent opening files, 4.87\% on closing them, 40.76\% on reading files, 8.28\% on writing, and 3.05\% on sleep operations.
Despite over 40\% of cycles being consumed by file reads, only 0.06\% of the total execution time is attributed to the \texttt{native\_msr} function. From these results, we conclude that: (1) approximately 55\% of the runtime could potentially be optimized by \refuser, and (2) if file read operations are minimized and optimized, up to 94.6\% of the total execution time could become available for \refkernel{}.

In Figure~\ref{fig:flameUser}, a significant shift in performance characteristics is evident. \refuser spends 69.6\% of its cycles on file reads, 1.20\% on file writes, and 24.35\% on sleep. File open and close operations occur so rarely that they are not captured in the measurement. Approximately 3.72\% of the total execution time is now attributed to the \texttt{native\_msr} function. The substantial increase in time spent on file reads and sleep is likely due to other operations becoming so infrequent that their relative share of compute time increases. These differences, together with the reduction in time spent on non-critical operations (shown in red), suggest that \refuser uses less than 50\% of the CPU cycles required by PowerJoular.




The flame graph in Figure~\ref{fig:flameKernel} aligns precisely with our expectations. \refkernel spends 14.68\% of its cycles on file reads, 54.35\% on file writes, and 28.59\% on sleep. The \texttt{native\_msr} function now executes in a separate kernel task and is therefore not included in this measurement. Given that both \refuser and \refkernel perform the same number of write operations during execution, it is reasonable to assume that the read and sleep operations here have a similar cycle cost to file writes in \refuser. Consequently, we estimate that \refkernel consumes roughly 5\% of the compute time required by \refuser, excluding any additional overhead introduced in kernel space. This represents an approximate 97.25\% performance improvement over PowerJoular, not accounting for the cost of the kernel process.

Furthermore, the extensive call stack of the read function in Figure~\ref{fig:flameUser} influenced the design decisions behind \refkernel. In the standard Linux MSR module, each MSR access triggers memory allocation at call time and performs multiple safety checks to ensure register availability. In contrast, \refkernel verifies safety and allocates memory for the four RAPL MSRs during module insertion. This design allows us to invoke the \texttt{native\_msr} function directly, minimizing the call chain. Such an optimization is feasible because \refkernel provides a specialized interface restricted to specific MSRs.
\vspace{-.2cm}
\section{Study Design}
This study investigates the overhead, namely the \textit{extra time and energy a measurement tool consumes beyond the minimum required purely for observation}.
We examine overhead from two sources: (1) the implementation of a tool, which varies in efficiency based on code design and inclusion of non-essential features, and (2) the function calls used to retrieve RAPL energy readings.
In this work, we focus on measurements taken at 1 kHz, that is, one sample per millisecond, which corresponds to the theoretical maximum sampling rate supported by RAPL \cite{hahnel2012measuring}.
We also hypothesize that using a tool with a basic implementation may reduce possible overhead due to the fewer function calls required to expose RAPL energy values. 
Based on this objective, we formulate the following research questions:

\RQone
For RQ1, we design a controlled experiment where a set of widely recognized energy measurements tools are compared against our two custom tools, described in Section \ref{sec:background}.
Overhead is measured by comparing the execution time and energy consumption of a benchmark when using a tool against a baseline recorded without any tools.
To obtain the baseline, we wrap the benchmark execution with a single RAPL function call before and after the run, without involving any additional tools.

\RQtwo

In RQ2, we examine the function calls (as detailed in Section \ref{sec:background}) that RAPL-based tools invoke when transitioning from user space to kernel space during measurements.
Our goal is to identify the calls that contribute most to execution time.
To achieve this, we profile these calls to determine which are the slowest and may therefore be optimized in future work.
For this purpose, we develop a kernel module to profile the calls during the execution of multiple benchmarks called \refbench.

\subsection{Subject Selection}
\label{sec:subject}
To answer RQ1, we select a set of popular tools for energy measurements and compare them against to \refkernel and \refuser.
Our selection is driven by \textit{popularity} and \textit{relevance} of the tool in the software energy optimization community.
We quantify popularity as GitHub stars, which should be above 1000, and relevance by mentions in scientific literature.
The collection of the tools led to a total of eight subjects: Perf \cite{perf}, Turbostat \cite{turbostat}, PowerJoular \cite{noureddine-ie-2022}, CodeCarbon \cite{codecarbon}, Scaphandre \cite{codecarbon}, EnergiBridge \cite{sallou2023energibridge}, Powerletrics \cite{geerd2025powerletrics}, and EcoFloc \cite{ecofloc}.
Some of the analyzed tools are unable to increase the sampling rate to 1 kHz.
In 3 such cases, we manually patched them.
Among the tools reviewed, PowerJoular, CodeCarbon, and Scaphandre required minor patches (fewer than 10 lines of code each) to support sampling rates below 1 Hz.
For PowerJoular, we substituted the fixed one-second delay with a one-millisecond delay.
For Scaphandre and CodeCarbon, we altered the parameter type used to define the polling interval from an integer to a floating-point number.
The tools otherwise accommodate higher rates.
We were not able to include EcoFloc, PowerLetrics, and EnergiBridge, as we experienced crashes at high polling rates or they had not yet reached a level of maturity to give accurate results at the time of the experiment. We chose to include the broadest possible set of tools, even when patching was required, to maximize the heterogeneity of our subject set. We hypothesize that overhead does not stem solely from the RAPL access mechanism, but also from features unrelated to polling energy values from RAPL, such as logging, reporting, or process monitoring, as well as from the quality of their implementation. Studying a diverse range of tools is necessary to determine whether such factors introduce significant overhead.
%
%
\uline{That said, it still is worth noting that none of these tools are meant to sample RAPL at 1 kHz (except arguably perf)} \cite{raffin2024dissecting}.
This paper applies the tested tools to a use case for which they were neither designed nor validated. 
Therefore, \textit{poor performance in this context does not reflect the effectiveness of the tools within their intended applications}.

For RQ2, we compare the execution time of eleven functions.
Four of them are used to directly read energy values from MSRs: \texttt{rdmsr 0x611}, \texttt{rdmsr 0x639}, \texttt{rdmsr 0x641}, and \texttt{rdmsr 0x619}.
Another four functions are "potentially supportive functions", that is, functions that assist or enable RAPL reading indirectly rather than performing the core data access themselves.
The instruction \texttt{cpuid 0x1 0} is a standard serializing instruction, and comparing the performance variations of different \texttt{rdmsr} instructions against it will help us determine whether they have a similar performance impact \cite{fog_instruction_tables}.
\texttt{rdmsr 0x19C} accesses a CPU register that reports temperature and power-throttling information. Although it touches on RAPL-related behavior (like reduced frequency due to power limits), it is not RAPL-specific, so reading it may have different overhead or timing effects than reading the actual RAPL registers \cite{intel_sdm}.
\texttt{proc\_read} handles \texttt{\/proc} interfaces (e.g., CPU info), and \texttt{sys\_read} supports sysfs paths like \texttt{/sys/class/powercap}, both of which are common in high-level RAPL access \cite{zhang2021red}.
We include three instructions irrelevant to RAPL \texttt{cpuid 0x10} \texttt{mov eax ebx} and \texttt{nop}.
These represent typical generic operations on modern x86 CPUs and serve as a baseline for non-RAPL instructions.
By contrasting RAPL functions against sub-single-cycle instructions, we isolate the overhead from RAPL access itself. Subtracting the baseline loop cost from the RAPL loop cost removes iteration overhead, while keeping the margin of underestimation minimal.

\subsection{Experiment Variables}
We design two experiments, each with different \textit{independent} and \textit{dependent} variables.
For RQ1, the independent variables are the \textit{benchmarks} profiled by each subject (i.e., a tool).
These functions are drawn from the NAS Parallel Benchmarks: \textit{bt} solves 3D compressible Navier-Stokes equations (compute-intensive, memory-bound), \textit{cg} performs conjugate gradient eigenvalue computation for a large sparse symmetric matrix (irregular memory access, communication-heavy), \textit{ft} benchmarks 3D fast Fourier transforms for solving partial differential equations (floating-point intensive), \textit{mg} tests multigrid hierarchical memory and computation (recursive), \textit{ep} measures parallel floating-point kernels like random number generation (highly parallelizable compute), and \textit{is} implements integer sort for particle method simulations (compute intensive).
The dependent variable is the profiling time, namely the time required by a tool to complete the measurement of a benchmark execution. 
Additionally, we compute the absolute and relative \textit{overhead}. 
The former is calculated by subtracting the median energy or execution time measured without any tool from the corresponding median obtained using a tool, while the relative overhead is normalized using the baseline and expressed in percentage.
%
For RQ2, we profile the instructions described in Section \ref{sec:subject}, the dependent variable is the execution time of each instruction in milliseconds.

\subsection{Experiment Design}
The experiment follows a \textit{full factorial} design.  
For RQ1, our samples consist of tuples composed of a tool, the corresponding subject, and a benchmark.  
The set of tools includes the subjects described in Section \ref{sec:subject}, along with the baseline representing measurements performed without any tool.  
We use six benchmarks from the NAS suite: bt, cg, ft, mg, ep, and is.
Each treatment is executed 15 times in random order to ensure consistent behavior across runs and to mitigate potential effects introduced by specific tools or benchmarks \cite{wohlin2012experimentation}.  
In total, we perform $8 \text{ tools} \times 6 \text{ benchmarks} \times 15 \text{ repetitions} = 720 \text{ runs}$.
The experiment for RQ2 uses the instructions described in Section \ref{sec:subject} as subjects.
We repeat the execution of a single operation 100,000 times in a loop for each benchmark.
In this setup, a run profiles a batch of 100,000 executions of an instruction.
Aggregating a large number of iterations increases the measurement resolution of these instructions and helps reveal potential differences in their execution times.
Because these instructions execute very quickly, profiling a single instance can be challenging.
This way of benchmarking x86/64 instructions resembles the approach used to make Agner Fog's instruction tables and the uops.info database, both of which rely on high-iteration microbenchmarks to profile fast x86 instructions \cite{fog_instruction_tables, uops_info}. 
We use a for loop of instructions and subtract the cost of a loop of the fastest instructions because that suits the needs and level of detail required. 
Each batch of 100,000 executions is repeated 15 times in random order, interleaved with the benchmarks from RQ1, with the same 30-second sleep before and after. 
We run a total of $11 \text{ instructions} \times 15 \text{ repetitions} = 165 \text{ runs}$.

\subsection{Data Analysis}
\label{sec:data}
Data analysis was conducted following the procedure described by Wohlin et al.~\cite{wohlin2012experimentation}. 
For RQ1, the groups comprise time and energy measurements obtained by combining different tools and benchmarks. 
Each tool–benchmark pair consists of a set of 15 time and energy values, where each value represents the time or energy consumed during a single run. 
Similarly, for RQ2, each group corresponds to the 15 values collected per instruction.
%
The experimental data are analyzed using descriptive statistics and visualizations.  
We assess the normality of each group using the Shapiro–Wilk test.
Due to the non-normality of our results, we use Kruskal–Wallis to test our hypothesis.
To explore pairwise differences between groups, we perform a post hoc analyses using Dunn's test with the Bonferroni correction to control the family-wise error rate and reduce the likelihood of false positives.
We quantify the magnitude of observed differences by calculating effect sizes using Cliff's delta.

\section{Experiment Execution}
\begin{figure}
  \includegraphics[width=\linewidth]{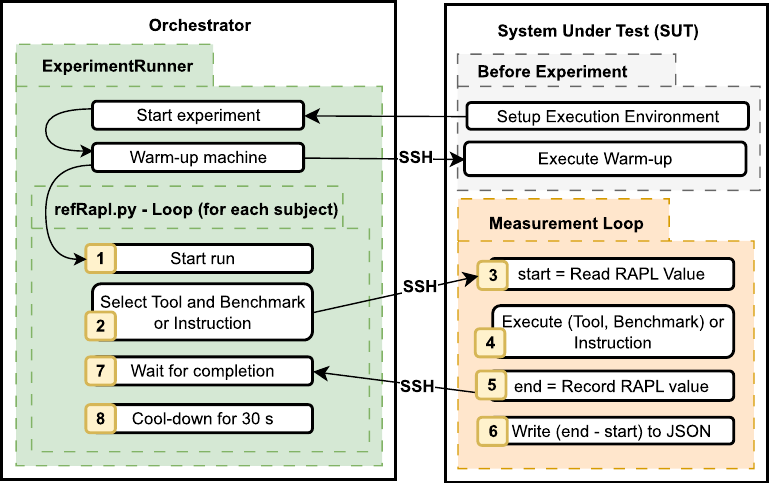}
  \caption{Runner and System Under Test Experimental Steps}
  \label{fig:expDiag}
\end{figure}
This section details the experimental preparation, setup, and execution.  
Figure \ref{fig:expDiag} illustrates the experimental environment, which consists of two machines: an \textit{orchestrator} and the \textit{system under test} (SUT).
The orchestrator defines and manages the experimental runs by specifying combinations of benchmarks and tools.  
Communication between the two machines occurs over SSH, enabling the exchange of commands and files.
To ensure unbiased results and minimize the influence of external or temporal factors, the orchestrator randomizes the order of runs.
We conduct two 30-second warm-up tests: one using CodeCarbon with the mg benchmark and one using Scaphandre with the bt benchmark, selected in random order.
After the warm-up phase, the experiments begin.
A 30-second cooldown period is included between runs to allow the system to stabilize and prevent interference from residual load or thermal effects \cite{apsan2025generating}.
%
After the warm-up, \texttt{refRapl.py} (Figure \ref{fig:expDiag}) runs an automated cycle where the orchestrator:
(i) selects a benchmark–tool pair; 
(ii) sends an SSH command to start it on the SUT; 
(iii) the SUT records the initial RAPL reading; 
(iv) executes the benchmark; 
(v) takes the final reading; 
(vi) stores results to JSON; and 
(vii) cools down for 30 s.
All the scripts used to orchestrate the experiment can be found in the \texttt{experiment-runner} folder of our replication package, along with the measurements in the \texttt{refrapl\_data} folder \cite{ReplicationPackage}.
%
In summary, each experiment wraps the execution of a benchmark–tool combination between two RAPL measurements to determine overhead in time and energy caused by the tool. For baseline measurements (i.e., without tools), only the benchmark is executed, still framed by RAPL readings for consistency.

\subsection{Experimental Setting}
The orchestrator machine is a Lenovo ThinkPad x390 with an Intel 8365U running between 1.6 and 4.1 GHz, an integrated GPU, 16GB of RAM, and 256Gb of solid-state storage. It was running NixOS 24.04 at the time of the test. In practice, the orchestrator could be any machine that can install the following requirements.
Before the execution of the experiment, it is set up with Python3 and SSH.
The tool used to orchestrate the experiment is ExperimentRunner \cite{karsten2025experiment}, a tool written in Python that facilitate the definition of all the steps of an experiment, including the number of runs, their order, and the collection of the metrics.
Currently, the tool comes with plugins to integrate well-known profiles (e.g., EnergiBridge, PowerJoular) into the experiment flow.
The replication package includes two files, requirements.txt and shell.nix, to set up everything the machine needs to run the experiment properly. These include numpy, pandas, and tabulate for data processing, psutil for process management, and pickle for compression and data transfer. 

The SUT consists of an Intel NUC8i7HVK running NixOS~24.04. 
We chose this device because it allows tracking all four RAPL domains, including memory.
It has an Intel 8809G CPU at 3.1 to 4.2 GHz, 64GB of RAM, 1TB of storage, and an AMD Radeon RX Vega M GH Graphics. 
NixOS enables reproducible, machine-independent configuration and simplifies the development and deployment of kernel modules. In principle, these results can be achieved on any Linux distribution, but our automation, build, and execution scripts rely on Nix commands and conventions. 
The key configuration details are to enable the SSH server with key-based authentication, install both Perf and Turbostat, and ensure the required kernel modules for RAPL are available at boot time. 
Before running any experiments, \textit{all components and tools included in the replication package must be compiled} \cite{ReplicationPackage}. 
This includes both user-level tools and kernel modules. 
The tools used as subjects, which we do not supply source code for, must be installed at the system level. 
As described in Section~\ref{sec:subject}, we patch some of these tools to enable measurements at 1\,kHz. 
Our replication package contains forks of the modified tools (CodeCarbon, Scaphandre, and PowerJoular) along with a \texttt{shell.nix} file to facilitate their compilation. In the case of CodeCarbon, the tool is fully interpreted and the \texttt{shell.nix} provides a runtime environment.
Additionally, the two kernel modules used to read RAPL values (\refkernel) and benchmark instructions for RQ2 (\refbench) must also be compiled. 
This can be done via the nix build scripts located in their respective directories (i.e, \texttt{raplmod} and \texttt{microbench}). 
Finally, the replication package also contains a copy of the NAS benchmarks used in the experiments, together with a Makefile for their compilation and a \texttt{shell.nix} with the gfortran compiler.

\section{Results}
\label{sec:Results}
This section presents the results of our study for each research question. For RQ1, the analysis aims to determine whether using tools at 1 kHz introduces noticeable differences in execution time and, consequently, any overhead compared to executions without tools. The objective of RQ2 is to measure the execution time of the instructions involved in the RAPL call chain, from user space to the kernel, and compare it to that of simple instructions (e.g., mov, nop) to assess the extent to which the former may delay the retrieval of energy values.
For brevity, this section reports only time overhead results. The replication package includes energy overhead data and the full set of measurements and statistical test code \cite{ReplicationPackage}.

\RQone

\begin{table*}
\centering
\footnotesize
\caption{Execution time statistics (in seconds) for the NAS Parallel Benchmarks. Green/red cells indicate the lowest/highest relative overhead (\%$\Delta$); highest median (50\%) per benchmark in bold. Abbreviations: N\_T: No\_Tools, R\_U: \refuser, R\_K: \refkernel, Perf: Linux perf, PJ: PowerJoular, Tur: Turbostat, Sca: Scaphandre, CC: CodeCarbon.}
\setlength{\tabcolsep}{1.2pt}
\begin{tabular}{l*{3}{*{8}{r}}}
\toprule
          & \multicolumn{8}{c}{\textbf{bt}} 
          & \multicolumn{8}{c}{\textbf{cg}} 
          & \multicolumn{8}{c}{\textbf{ep}} \\
\cmidrule(lr){2-9}\cmidrule(lr){10-17}\cmidrule(lr){18-25}
\textbf{Statistic} &
\textbf{No\_T} & \textbf{R\_K} & \textbf{R\_U} & \textbf{Perf} & \textbf{PJ} & \textbf{Tur} & \textbf{Sca} & \textbf{CC} &
\textbf{No\_T} & \textbf{R\_K} & \textbf{R\_U} & \textbf{Perf} & \textbf{PJ} & \textbf{Tur} & \textbf{Sca} & \textbf{CC} &
\textbf{No\_T} & \textbf{R\_K} & \textbf{R\_U} & \textbf{Perf} & \textbf{PJ} & \textbf{Tur} & \textbf{Sca} & \textbf{CC} \\
\midrule
\textbf{mean} &
196.21 & 196.34 & 195.54 & 197.00 & 199.48 & 209.83 & 218.10 & 228.70 &
43.68 & 43.61 & 43.00 & 43.99 & 45.15 & 47.52 & 51.33 & 50.22 &
268.21 & 282.25 & 272.11 & 277.64 & 289.52 & 305.16 & 340.78 & 389.73 \\
\textbf{std} &
0.20 & 0.35 & 1.12 & 0.21 & 0.21 & 0.46 & 0.59 & 0.44 &
0.13 & 0.09 & 0.80 & 0.10 & 0.11 & 0.07 & 0.22 & 0.84 &
5.45 & 21.11 & 7.06 & 3.81 & 3.21 & 4.91 & 7.59 & 3.28 \\
\textbf{min} &
195.92 & 195.66 & 194.41 & 196.78 & 199.19 & 209.08 & 217.22 & 227.70 &
43.50 & 43.46 & 42.12 & 43.87 & 44.99 & 47.39 & 51.02 & 47.38 &
264.22 & 265.23 & 262.84 & 276.02 & 286.27 & 302.41 & 329.06 & 386.08 \\
\textbf{50\%} &
196.21 & 196.29 & 194.84 & 196.95 & 199.48 & 209.88 & 218.10 & \textbf{228.79} &
43.67 & 43.60 & 43.58 & 43.97 & 45.12 & 47.50 & \textbf{51.36} & 50.47 &
265.28 & 267.90 & 273.05 & 276.59 & 288.84 & 303.39 & 340.56 & \textbf{389.29} \\
\textbf{max} &
196.58 & 196.86 & 197.45 & 197.60 & 200.00 & 210.34 & 219.26 & 229.31 &
43.89 & 43.77 & 43.95 & 44.26 & 45.35 & 47.71 & 51.78 & 50.77 &
277.85 & 325.22 & 284.02 & 291.20 & 298.15 & 317.29 & 356.04 & 398.52 \\
\textbf{\bm{$\Delta t$}} &
0.00 & 0.08 & -1.37 & 0.74 & 3.28 & 13.67 & 21.89 & 32.58 &
0.00 & -0.07 & -0.09 & 0.30 & 1.45 & 3.83 & 7.69 & 6.80 &
0.00 & 2.62 & 7.77 & 11.31 & 23.56 & 38.11 & 75.28 & 124.01 \\
\textbf{\bm{$\%\Delta$}} &
0.00 & 0.04 & \cellcolor{green!20}-0.70 & 0.38 & 1.67 & 6.97 & 11.16 & \cellcolor{red!20}16.60 &
0.00 & -0.17 & \cellcolor{green!20}-0.20 & 0.69 & 3.31 & 8.77 & \cellcolor{red!20}17.60 & 15.57 &
0.00 & 0.99 & 2.93 & 4.26 & 8.88 & 14.37 & 28.38 & \cellcolor{red!20}46.75 \\
\midrule
          & \multicolumn{8}{c}{\textbf{ft}} 
          & \multicolumn{8}{c}{\textbf{is}} 
          & \multicolumn{8}{c}{\textbf{mg}} \\
\cmidrule(lr){2-9}\cmidrule(lr){10-17}\cmidrule(lr){18-25}
&
\textbf{No\_T} & \textbf{R\_K} & \textbf{R\_U} & \textbf{Perf} & \textbf{PJ} & \textbf{Tur} & \textbf{Sca} & \textbf{CC} &
\textbf{No\_T} & \textbf{R\_K} & \textbf{R\_U} & \textbf{Perf} & \textbf{PJ} & \textbf{Tur} & \textbf{Sca} & \textbf{CC} &
\textbf{No\_T} & \textbf{R\_K} & \textbf{R\_U} & \textbf{Perf} & \textbf{PJ} & \textbf{Tur} & \textbf{Sca} & \textbf{CC} \\
\midrule
\textbf{mean} &
54.84 & 54.85 & 54.36 & 54.32 & 54.38 & 57.11 & 56.95 & 59.80 &
102.50 & 102.56 & 102.15 & 103.57 & 104.44 & 109.62 & 107.45 & 109.05 &
25.41 & 25.41 & 24.96 & 25.49 & 25.61 & 26.10 & 26.53 & 26.68 \\
\textbf{std} &
0.30 & 0.29 & 0.65 & 0.40 & 0.33 & 0.14 & 0.31 & 0.48 &
0.05 & 0.09 & 0.85 & 0.06 & 0.10 & 0.17 & 0.30 & 0.13 &
0.03 & 0.03 & 0.69 & 0.03 & 0.03 & 0.03 & 0.06 & 0.24 \\
\textbf{min} &
54.25 & 54.08 & 53.39 & 53.84 & 53.92 & 56.90 & 56.36 & 58.63 &
102.41 & 102.43 & 101.10 & 103.48 & 104.20 & 109.34 & 106.92 & 108.87 &
25.38 & 25.38 & 24.00 & 25.45 & 25.56 & 26.04 & 26.46 & 25.97 \\
\textbf{50\%} &
54.80 & 54.94 & 54.74 & 54.26 & 54.33 & 57.08 & 56.89 & \textbf{59.91} &
102.51 & 102.54 & 102.79 & 103.55 & 104.43 & \textbf{109.62} & 107.44 & 109.04 &
25.41 & 25.41 & 25.41 & 25.47 & 25.62 & 26.10 & 26.51 & \textbf{26.78} \\
\textbf{max} &
55.39 & 55.15 & 55.07 & 55.30 & 55.16 & 57.49 & 57.44 & 60.51 &
102.56 & 102.72 & 102.90 & 103.75 & 104.57 & 109.91 & 107.94 & 109.31 &
25.48 & 25.47 & 25.49 & 25.58 & 25.65 & 26.17 & 26.66 & 26.82 \\
\textbf{\bm{$\Delta t$}} &
0.00 & 0.14 & -0.06 & -0.55 & -0.47 & 2.28 & 2.09 & 5.11 &
0.00 & 0.03 & 0.28 & 1.04 & 1.92 & 7.11 & 4.93 & 6.53 &
0.00 & 0.00 & 0.00 & 0.06 & 0.21 & 0.69 & 1.10 & 1.37 \\
\textbf{\bm{$\%\Delta$}} &
0.00 & 0.26 & \cellcolor{green!20}-0.11 & \cellcolor{green!20}-1.00 & -0.87 & 4.15 & 3.81 & \cellcolor{red!20}9.32 &
0.00 & 0.03 & 0.27 & 1.02 & 1.87 & 6.93 & 4.81 & \cellcolor{red!20}6.37 &
0.00 & 0.00 & \cellcolor{green!20}-0.04 & 0.25 & 0.82 & 2.73 & 4.34 & \cellcolor{red!20}5.38 \\
\bottomrule
\end{tabular}
\label{tab:time_stats}
\end{table*}

\begin{figure}
    \centering
    \includegraphics[width=\linewidth]{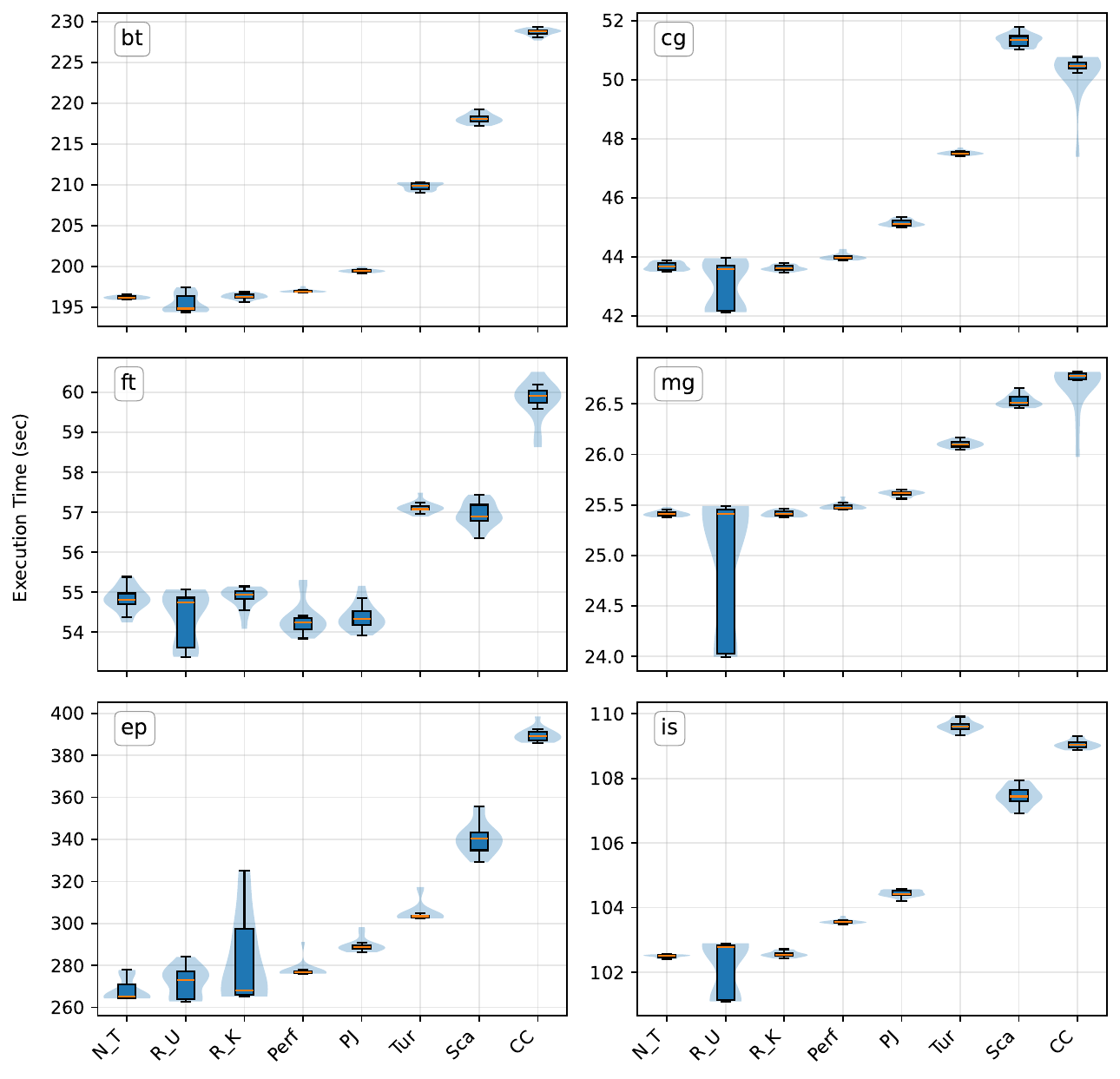}
    \caption{Profiling time (in seconds) for each tool across benchmarks. Abbreviations: N\_T: No\_Tools, R\_U: \refuser, R\_K: \refkernel, Perf: Linux perf, PJ: PowerJoular, Tur: Turbostat, Sca: Scaphandre, CC: CodeCarbon.}
    \label{fig:profiling_time}
    \vspace{-.6cm}
\end{figure}

Table \ref{tab:time_stats} presents descriptive statistics of execution times across six NAS benchmarks, bt, cg, ep, ft, is, and mg, for our eight subjects: No\_Tools (N\_T), \refuser (R\_U), \refkernel (R\_K), Perf, PowerJoular (PJ), Turbostat (Tur), Scaphandre (Sca), CodeCarbon (CC).
Color highlighting indicates the lowest and highest median values per benchmark, excluding No\_Tools, while the highest median is shown in bold.
The symbols $\Delta t$ and $\%\Delta t$ represent the absolute and the relative overhead, respectively.
Results show that our tools (R\_K and R\_U) maintain a profiling time close to the No\_T baseline across all NAS Parallel Benchmarks, with R\_U frequently achieving the lowest relative overhead.
For example, R\_U yielded negative overheads of -0.70\%, -0.20\%, and -0.11\% for bt, cg, and ft, respectively, indicating slight improvements over the baseline.
In contrast, commercial tools such as Sca and CC exhibit the highest slowdowns, reaching up to +17.6\% for cg and +46.8\% for ep.
The absolute overheads ($\Delta t$) confirm this trend, showing deviations under ten seconds for R\_K and R\_U compared to delays of several tens of seconds under the heaviest tools. 
Considering the medians (50\%), R\_U and R\_K are tightly around the baseline, suggesting consistent and predictable overhead behavior.
Standard deviations (std) remain low.
It is typically $<$1 for most benchmarks, except for ep, where variability is consistently higher across tools.
These results are further supported by the box plots in Figure \ref{fig:profiling_time}.
The figure shows that, in some cases, the profiling times of Tur and Sca exceed those of CC, although CC remains the slowest in the remaining cases. In particular, the median profiling time of Sca is higher than that of CC for the cg benchmark, whereas Tur shows a higher median profiling time for the is benchmark.
Overall, \textit{our solutions R\_K and R\_U incur the least overhead, while Sca and CC consistently exhibit the highest}.

\begin{table}
\centering
\scriptsize
\caption{Dunn's post-hoc test $p$-values (Cliff's $\delta$ effect sizes in parentheses) for \textit{No\_Tools} (N\_T) vs.\ other tools. Abbreviations: 
\textbf{R\_U}: \refuser, \textbf{R\_K}: \refkernel, \textbf{Perf}: Linux perf, \textbf{PJ}: PowerJoular, \textbf{Tur}: Turbostat, \textbf{Sca}: Scaphandre, \textbf{CC}: CodeCarbon.}
\setlength{\tabcolsep}{3pt}
\label{tab:dunn-notools}
\begin{tabular}{>{\raggedright\arraybackslash}p{0.4cm} *{6}{>{\centering\arraybackslash}p{1.15cm}}}
\toprule
\multirow{2}{*}{} & \multicolumn{6}{c@{}}{\textbf{Benchmark}} \\
\cmidrule(lr){2-7}
 & \textbf{bt} & \textbf{cg} & \textbf{ep} & \textbf{ft} & \textbf{is} & \textbf{mg} \\
\midrule
\textbf{R\_K}  & \cellcolor{small}1.000 (-0.28) & \cellcolor{small}1.000 (0.25) & \cellcolor{large}1.000 (-0.54) & \cellcolor{small}1.000 (-0.16) & \cellcolor{medium}1.000 (-0.42) & 1.000 (0.04) \\
\textbf{R\_U}  & \cellcolor{small}1.000 (0.32) & \cellcolor{medium}1.000 (0.35) & \cellcolor{small}1.000 (-0.16) & \cellcolor{medium}1.000 (0.37) & \cellcolor{small}1.000 (-0.20) & 1.000 (0.06) \\
\textbf{Perf}  & \cellcolor{large}0.915 (-1.00) & \cellcolor{large}1.000 (-0.99) & \cellcolor{large}1.000 (-0.70) & \cellcolor{large}1.000 (0.70) & \cellcolor{large}0.178 (-1.00) & \cellcolor{large}0.939 (-0.93) \\
\textbf{PJ}    & \cellcolor{large}\textbf{0.012} (-1.00) & \cellcolor{large}\textbf{0.041} (-1.00) & \cellcolor{large}\textbf{0.025} (-1.00) & \cellcolor{large}1.000 (0.71) & \cellcolor{large}\textbf{0.003} (-1.00) & \cellcolor{large}\textbf{0.017} (-1.00) \\
\textbf{Tur}   & \cellcolor{large}\textbf{<0.001} (-1.00) & \cellcolor{large}\textbf{<0.001} (-1.00) & \cellcolor{large}\textbf{<0.001} (-1.00) & \cellcolor{large}\textbf{0.019} (-1.00) & \cellcolor{large}\textbf{<0.001} (-1.00) & \cellcolor{large}\textbf{<0.001} (-1.00) \\
\textbf{Sca} & \cellcolor{large}\textbf{<0.001} (-1.00) & \cellcolor{large}\textbf{<0.001} (-1.00) & \cellcolor{large}\textbf{<0.001} (-1.00) & \cellcolor{large}0.086 (-1.00) & \cellcolor{large}\textbf{<0.001} (-1.00) & \cellcolor{large}\textbf{<0.001} (-1.00) \\
\textbf{CC}    & \cellcolor{large}\textbf{<0.001} (-1.00) & \cellcolor{large}\textbf{<0.001} (-1.00) & \cellcolor{large}\textbf{<0.001} (-1.00) & \cellcolor{large}\textbf{<0.001} (-1.00) & \cellcolor{large}\textbf{<0.001} (-1.00) & \cellcolor{large}\textbf{<0.001} (-1.00) \\
\bottomrule
\multicolumn{7}{l}{\textit{P-values < 0.05 are in \textbf{bold}. Effect sizes: \marktext{large}{Large} - \marktext{medium}{Medium} - \marktext{small}{Small} - \marktext{white}{Negligible}}} \\
\end{tabular}
\vspace{-.6cm}
\end{table}

We test the normality of each group using the Shapiro–Wilk test.
As mentioned in Section \ref{sec:data}, a group is defined as a combination of tool and benchmark.
The results are mixed: some groups followed a normal distribution, while others did not. 
Consequently, we applied the Kruskal–Wallis test to compare the groups.
This test determines whether there are significant differences across the tools for the same benchmark.
Thus, the test yields six values that are all lower than our significance level of $0.05$, indicating differences among the tools' profiling time across benchmarks.
We conduct a post hoc analysis using Dunn's test with Bonferroni correction and quantify the magnitude of the differences using Cliff's delta.
Table \ref{tab:dunn-notools} presents the results of the post-hoc analysis comparing \textit{No\_Tools} (N\_T), i.e., the baseline, against the tools across the six benchmarks. 
Each cell reports the p-value from the Dunn test and the Cliff's delta ($\delta$) effect size in parentheses, indicating the magnitude and direction of the difference. 
Smaller p-values (in bold) denote statistically significant contrasts, while the sign and size of $\delta$ describe whether the difference favors No\_Tools (negative $\delta$) or the compared tool (positive $\delta$).
Three main patterns emerge across benchmarks.
We notice non-significant differences comparing R\_K and R\_U with the baseline.
Both tools show p = 1.000 in all benchmarks, meaning there are no statistically significant differences compared to No\_Tools.
Effect sizes are small to medium in magnitude (e.g., $\delta$ ranges from -0.42 to +0.37), suggesting minor differences across benchmarks.
We found mixed results for Perf and PJ.
Perf shows large effect sizes ($|\delta| \geq 0.70$) in all benchmarks but mostly non-significant p-values, except a marginal trend for is (p = 0.178).
In contrast, PJ shows significant differences in five out of six benchmarks (\(p = 0.012\text{–}0.041\)), all with \(\delta = -1.00\), indicating a very large negative effect of \textit{No\_Tools}. This result reflects a substantially higher execution time for PJ compared to \textit{No\_Tools}.
Finally, there are large, consistent differences for Tur, Scaph, and CC.
These tools all exhibit uniformly significant p-values (p > 0.05) across nearly every benchmark, paired with $\delta$ = -1.00, the maximum possible effect size in magnitude.
This means the difference from No\_Tools is both statistically significant and large.

In summary, the strongest effects appear for Tur, Scaph, and CC, each producing consistently significant and large differences (all p < 0.05, $\delta$ = -1.00).
Tools like PJ also show strong but less consistent significance.
R\_K and R\_U show negligible statistical evidence of difference, despite small to medium effect sizes.


\RQtwo

\begin{table*}[t]
\centering
\footnotesize
\caption{Execution time statistics (in milliseconds) for selected microbenchmarks. The lowest median (50\%) per row is highlighted in green, and the highest in red. The last row shows the median execution time per single execution (\(50\%\div100{,}000\)).}
\setlength{\tabcolsep}{3pt}
\begin{tabular}{l*{11}{r}}
\toprule
\textbf{Statistic} &
\textbf{cpuid 0x1 0} & 
\textbf{mov eax ebx} & 
\textbf{nop} & 
\textbf{rdmsr 0x611} & 
\textbf{rdmsr 0x639} & 
\textbf{rdmsr 0x641} & 
\textbf{rdmsr 0x619} & 
\textbf{rdmsr 0x19C} & 
\textbf{rdmsr 0x17} & 
\textbf{proc\_read} & 
\textbf{sys\_read} \\
\midrule
\textbf{mean} &
13.11 & 0.03 & 0.03 & 55.37 & 53.73 & 33.31 & 23.96 & 22.73 & 23.01 & 67.61 & 141.72 \\
\textbf{std} &
0.07 & 0.00 & 0.00 & 1.34 & 1.68 & 2.88 & 0.56 & 0.44 & 0.48 & 7.52 & 12.24 \\
\textbf{min} &
12.99 & 0.03 & 0.03 & 52.94 & 49.57 & 28.53 & 23.29 & 22.17 & 22.21 & 64.20 & 134.01 \\
\textbf{25\%} &
13.06 & 0.03 & 0.03 & 54.24 & 52.84 & 31.92 & 23.50 & 22.25 & 22.55 & 64.66 & 134.88 \\
\textbf{50\%} &
13.13 & 0.03 & \cellcolor{green!20}0.03 & 55.68 & 53.81 & 32.74 & 23.80 & 22.87 & 23.17 & 65.09 & \cellcolor{red!20}135.74 \\
\textbf{75\%} &
13.17 & 0.03 & 0.03 & 56.63 & 54.69 & 34.65 & 24.29 & 23.09 & 23.41 & 66.01 & 143.20 \\
\textbf{max} &
13.21 & 0.04 & 0.03 & 56.89 & 56.57 & 39.50 & 25.05 & 23.39 & 23.64 & 93.39 & 179.99 \\
\midrule
\textbf{50\%/100k} &
0.00013 & 3.00e$^{-7}$ & \cellcolor{green!20}3.00e$^{-7}$ & 0.00056 & 0.00054 & 0.00033 & 0.00024 & 0.00023 & 0.00023 & 0.00065 & \cellcolor{red!20}0.00136 \\
\bottomrule
\end{tabular}
\label{tab:microbenchmarks_ms}
\end{table*}

\begin{figure}
    \centering
    \includegraphics[width=\linewidth]{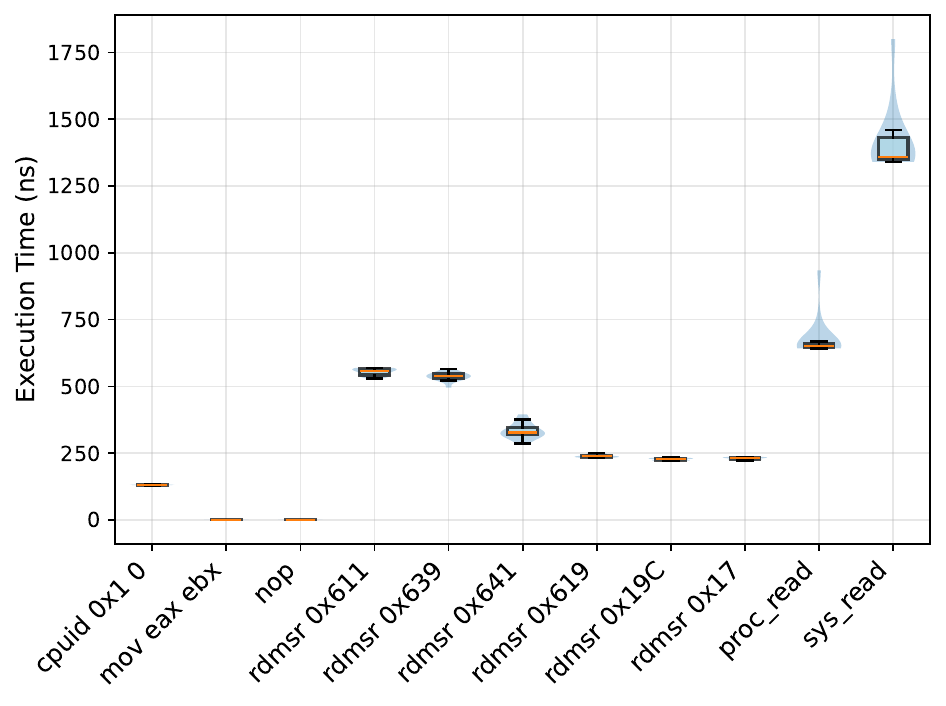}
    \caption{Execution time (in nanoseconds) per instruction execution. }
    \label{fig:instruction_time}
    \vspace{-.6cm}
\end{figure}

Table \ref{tab:microbenchmarks_ms} reports the execution time statistics (in milliseconds) for the chosen instructions, each measured over 100{,}000 iterations.
The last row presents the median value per batch normalized per single execution (50\% / 100{,}000).
As expected, the baseline instructions, \texttt{nop}, \texttt{mov eax, ebx}, and \texttt{cpuid 0x1 0}, show the lowest execution time overall. 
The \texttt{nop} and \texttt{mov} both show a median of 0.03\,ms per batch and 3$\times$10\textsuperscript{-7}\,ms per execution, the lowest among all tests (green-highlighted), while \texttt{cpuid} is considerably slower at a median of 13.13\,ms per batch (1.3$\times$10\textsuperscript{-4}\,ms per execution), roughly four orders of magnitude slower than trivial register moves.
The \texttt{rdmsr} instructions exhibit moderate latency, ranging from 23\,ms to 56\,ms per batch, depending on the specific register (e.g., \texttt{rdmsr 0x641} = 32.74\,ms, \texttt{rdmsr 0x639} = 53.81\,ms).
The system calls are the slowest overall.
In particular, \texttt{sys\_read} reaches a median of 135.74\,ms per batch (1.36$\times$10\textsuperscript{-3}\,ms per execution), about 10$\times$ slower than an \texttt{rdmsr} and over 4000$\times$ slower than \texttt{nop}.
The standard deviations are minimal for baseline instructions (\texttt{mov}, \texttt{nop}, \texttt{cpuid}), confirming stable timing for these low-level operations. 
In contrast, \texttt{rdmsr} variants show moderate variability (0.4-2.9\,ms), while system calls display substantially higher dispersion (7.5-12.2\,ms).
These results are confirmed by the box plot represented in Figure \ref{fig:instruction_time}. Simple CPU instructions like \texttt{cpuid}, \texttt{mov}, and \texttt{nop} show variations of less than 0.01\,ms. More complex instructions and system calls (\texttt{rdmsr}, \texttt{sys/proc\_read}) have higher std values, up to 12.24\,ms.
Overall, the table shows an expected performance hierarchy: \textit{user-space instructions execute in microseconds or less with minimal jitter, while system interactions introduce noticeable latency}.

The Shapiro-Wilk test shows that while some groups satisfy the assumption of normality, others do not.
Consequently, we use the Kruskal-Wallis test to examine potential differences in time values across instructional conditions.
Post-hoc comparisons are carried out using Dunn's test, and Cliff's delta is computed to quantify effect sizes.
The Kruskal-Wallis test reveals a statistically significant result ($H = 160.49$, $p = 2.56 \times 10^{-29}$), indicating substantial differences among groups.
These results are further supported by the outcomes of Dunn's test and the corresponding effect sizes.

Dunn's post-hoc test reveals clear clusters of instructions, with strong statistical significance ($p$ < 0.05) and predominantly large Cliff's $\delta$ effect sizes ($|\delta| \geq 0.58$).
\texttt{rdmsr 0x611/619/639/641} and syscalls show no significant intra-group differences ($p \geq 0.544$) but dominate simple instructions, namely \texttt{cpuid 0x1 0}, \texttt{mov eax ebx}, and \texttt{nop}.
The latter show non-significant differences among themselves ($p$ > 0.05, small/medium $\delta$).
These results confirm the trends observed in Table \ref{tab:microbenchmarks_ms} and Figure \ref{fig:instruction_time}, which show a clear distinction between simple instructions, the \texttt{rdmsr} instruction, and system calls (\texttt{proc/sys\_read}).
The latter exhibit significantly higher execution times compared to the assembly-level instructions.

\section{Discussion}
This paper demonstrates that RAPL-based tools can introduce overhead when profiling software systems at a high frequency.
We show that the overhead can come from the implementation complexity of the profile, as well as from the polling rate.
In fact, the latter can involve polling energy values from kernel to user space, and therefore involve some instructions that are notoriously slow in the process.
Our results align with \citet{raffin2024dissecting} in highlighting the efficiency benefits of simplicity. While their use of \texttt{perf} reduces latency by avoiding redundant system calls, our solutions (\refuser and \refkernel) achieve similar gains through the same principle, a finding we independently confirm. They further demonstrate the significant overhead of Scaphandre and CodeCarbon at 10 Hz.

\begin{figure}
    \centering
    \includegraphics[width=\linewidth]{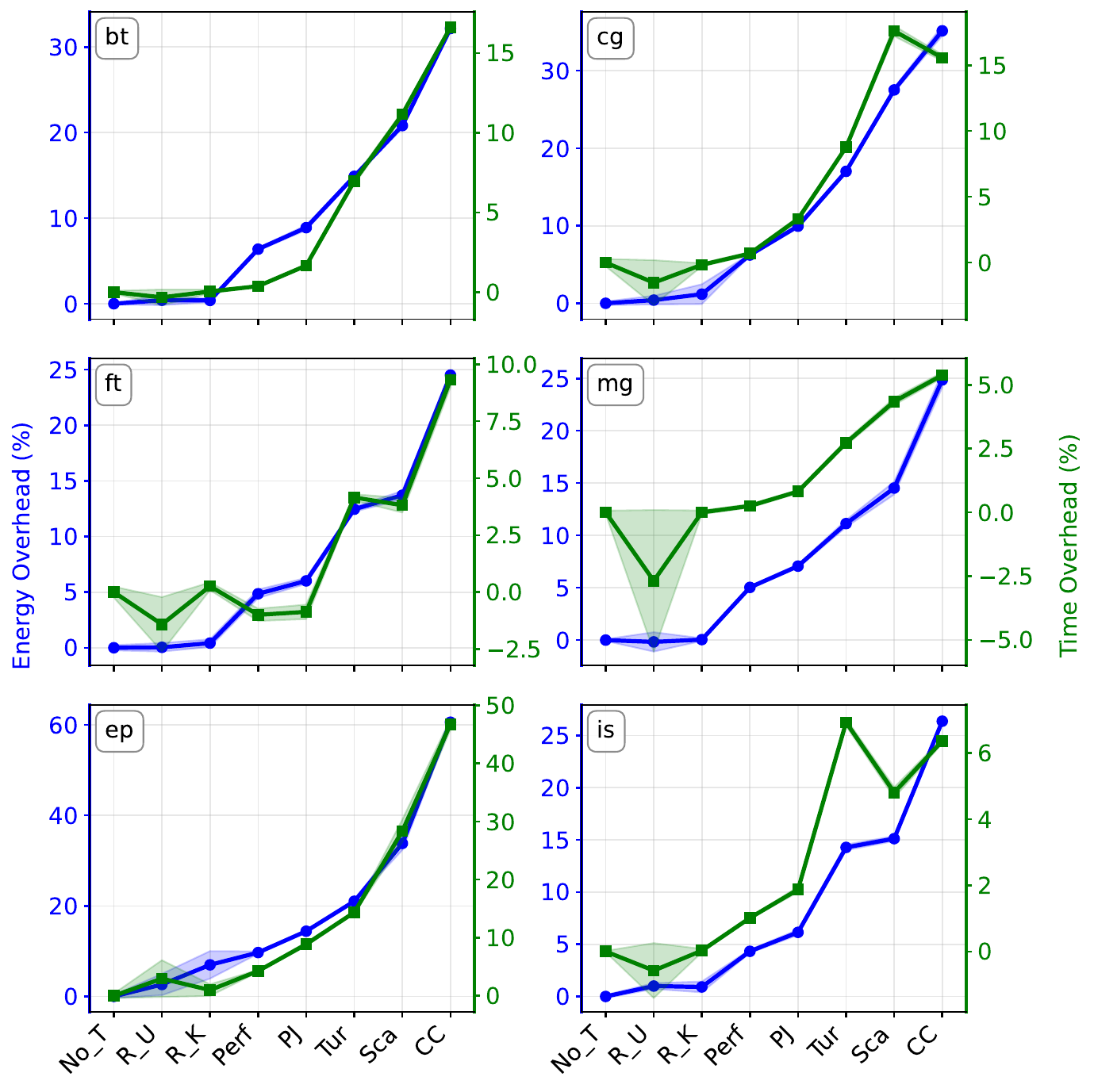}
    \caption{Lineplot of Energy and Time Relative Overhead for all NAS benchmarks \textbf{bt}, \textbf{cg}, \textbf{ep}, \textbf{ft}, \textbf{is}, and \textbf{mg} and all tools. Legend -  \textbf{No\_T}: No\_Tools, \textbf{R\_U}: \refuser, \textbf{R\_K}: \refkernel, \textbf{Perf}: Linux perf, \textbf{PJ}: PowerJoular, \textbf{Tur}: Turbostat, \textbf{Sca}: Scaphandre, \textbf{CC}: CodeCarbon.}
    \label{fig:rapl_box_all}
    \vspace{-0.6cm}
\end{figure}
Figure \ref{fig:rapl_box_all} shows the energy and time overhead \textit{relative} to the baseline for each subject in the study.
In all cases, already existing tools, Perf, PJ, Tur, Scaph, and CC, exhibit higher overhead compared to our solutions.
Notably, Scaph and CC consistently appear at the top of each graph, indicating a significant "observer effect" when reading energy values from RAPL at 1 kHz.
CC consistently presents the highest overhead, ranging from 5.38\% - 46.75\%.
This variability shows that the \textit{overhead may depend on the nature of the profiled code}.
Our approaches, \refuser and \refkernel, have a minimal impact on the measurement activity, keeping both execution time and energy consumption at their lowest levels.
We observe that \refkernel performs better in terms of time overhead but worse in terms of energy overhead compared to \refuser. Since ep is a highly parallel workload, it is possible that competition with another user-space process is less disruptive than contention with a kernel task.
Future work are needed to investigate these energy-time trade-offs.
In almost all scenarios, the green line (Time Overhead) and the blue line (Energy Overhead) move in parallel, showing the \textit{same trend across benchmarks time and energy overhead}.
In fact, besides the time overhead, we observe a significant energy overhead that can skew the outcomes, potentially leading to biased results in an empirical analysis. For instance, we measured an energy overhead exceeding 40\% when using CC profiling ep.

The results in Table \ref{tab:microbenchmarks_ms} indicate that calls to the proc and sys virtual file systems can take a substantial fraction of the execution time across an entire measurement. Remember that we are optimizing up to 10 file system access system calls per iteration. The system calls themselves could take upwards of 1 percent (0.01 ms) of the execution time. We are also measuring the most optimal RDMSR executions where the assembly instructions are directly inlined in the proc read functions (see \url{RefRAPL/microbenchmod/src/micro_bench.c} in the replication package). Recall from Section \ref{sec:flame} that the amount of time in a powercap sys read spent actually reading the MSR is 2.4\% of the syscall. Even the more optimal path taken by \refuser was still less than perfect. The use of powercap or the msr kernel module could be 10 times more overhead than the optimal inline version measured by microbench. Considered in terms of the execution budget of a CPU per millisecond, this data substantiates the idea that these functions could use up a substantial percentage of that budget in a millisecond.

An additional question emerges from the combination of results from RQ1 and RQ2. For RQ1, we show that \refuser and \refkernel are similar in performance, while in RQ2, we show that syscalls are a substantial source of overhead, and we know \refkernel optimizes out an additional four syscalls per RAPL read. Our best guess is that there is a natural cost of waking the CPU up or switching tasks every millisecond, and if the tool is substantially below that cost, it performs similarly regardless of how much work it does. This leads to the possibility that \refkernel represents an over-optimization. However, we argue that its improved developer ergonomics, potential for greater measurement accuracy, and demonstrated performance gains in certain workloads warrant further investigation.

\textit{Implications for Practitioners:} Profiling at very high frequencies (e.g., 1 kHz) can introduce significant overhead because frequent polling between kernel and user space is expensive. Practitioners should balance measurement granularity instead of defaulting to high-frequency sampling. Thus, \textit{they should opt for simpler tools (e.g., perf) for their experiments}. Simpler tools outperform more complex tools like Scaph and CC because they reduce system calls and internal processing. This suggests tool designers should aim for minimal, focused functionality to avoid the "observer effect" of their own instrumentation. Furthermore, accessing energy counters via proc or sysfs can consume a noticeable share of CPU time. Developers should be aware that frequent or unnecessary system-level reads can distort both timing and energy results. Finally, limit RAPL polling to only the domains you need. If your app is not memory or GPU intensive, measuring PP0 alone is sufficient.


\section{Threats to Validity}
This section discusses threats to validity and mitigation measures, categorized following Cook and Campbell \cite{Cook:1979}.

\textit{Internal Validity:} The experiment conducted for RQ1 evaluates seven tools: five commercial and two developed in-house.
Three of the commercial tools required patching, as they did not originally support measurements at a frequency of 1 kHz. 
This modification may have affected their functionality,\textit{ potentially introducing inefficiencies}.
However, the applied patches were minimal and straightforward.
For PowerJoular, we replaced a fixed one-second sleep interval with a one-millisecond interval.
For Scaphandre and CodeCarbon, we modified the argument type used to specify the polling period from an integer to a floating-point value.
After applying these changes, we manually verified that all tools continued to function correctly.
It is important to recall that the scope of our analysis \textit{is not a comparison of which tool performs best, but rather shows whether overhead can be introduced by profiling software energy usage at high frequency}.
Additionally, our tool selection may not be comprehensive of all commercial tools.
We mitigate this threat by choosing those most representative according to popularity and source code availability.
Similarly, for RQ2, only the most essential instructions (\texttt{nop}, \texttt{mov}, \texttt{syscall}, \texttt{rdmsr}) are considered, not the entire set of those present in the call chain.
%


\textit{External Validity:} 
External validity threats can come from the limited heterogeneity of the experimental setting and benchmarks.
Our main experiments ran on a single NixOS 24.11 machine (x86\_64 architecture), which risks hardware/OS-specific artifacts like kernel bugs, driver quirks, or thermal behaviors that may not generalize to other CPUs (e.g., ARM), GPUs, memory configs, or OSes.
We mitigated this by spot-checking all tools and benchmarks on two additional machines (varying NixOS versions/kernels) prior to main execution on NixOS 24.11, confirming robustness.
A full replication across architectures was infeasible due to resource constraints. Future work should include multi-platform execution.
Benchmarks came from a single source and toolchain, risking workload bias. 
We used the NAS suite for its heterogeneous, representative workloads.

\textit{Conclusion Validity:}
A source of conclusion validity can come from using tests not suited for the kind of collected data.
To mitigate this, we check the assumptions of the tests prior to execution and corroborate their results with visualizations.
Additionally, to further substantiate the basis of our hypotheses regarding the potential overhead introduced by measurement activities, we performed a CPU time allocation analysis using flame graphs, as described in Section \ref{sec:background}.
This analysis shows that the sources of overhead we aimed to optimize are present in another project but absent in \refuser and \refkernel.
%

\section{Related Work}

The accuracy of RAPL-based energy measurements has been extensively studied \cite{hahnel2012measuring, alt2024experimental, desrochers2016validation, ostapenco2024exploring}, though few works examine the associated flaws or overhead.  
Our work builds most directly on \citet{raffin2024dissecting}, who benchmarked RAPL access methods (MSR, powercap, perf-events, eBPF) across Intel and AMD CPUs using Rust-based tools. 
The paper finds perf-events optimal for low overhead despite similar measurement performance overall, highlighting pitfalls in tools like CodeCarbon or Scaphandre and recommending lightweight sampling up to ~1000 Hz. We adopt a similar approach, extending it with two custom solutions (\refuser and \refkernel) to investigate the causes of overhead.
Related studies acknowledge RAPL's impact at high sampling frequencies. \citet{van2025s} suggest keeping profilers lightweight, echoing our objectives. \citet{thamm2025strategies} propose practical approaches for integrating RAPL into workflow systems, reporting negligible overhead for short-lived tasks, while \citet{geerd2025powerletrics} introduce Powerletrics, an eBPF-based framework providing real-time per-process power metrics with minimal overhead.  Our study focuses not on tool development but on analyzing overhead sources in RAPL-based measurement tools at high sampling frequencies. 

The performance implications of system calls and instructions like \texttt{rdmsr} are well-documented \cite{herzog2021price, kuznetsov2022privbox, tanenbaum_modern_os}. System calls incur kernel traps and context switches, which can significantly increase overhead in high-frequency measurements. Prior work on HPC workloads \citet{miwa2023analyzing} confirm that system call latency remains non-negligible even on modern platforms. Instruction-level analyses by \citet{fog_instruction_tables} and \citet{uops_info} further show that \texttt{rdmsr} has roughly double the latency and half the throughput of \texttt{cpuid}, underscoring its cost in RAPL measurement scenarios.



\section{Conclusion}
This paper examines whether high-frequency energy measurement (1~kHz) with a RAPL-based tool introduces overhead. Such overhead may result from (i) tool complexity, including optional features like data aggregation, and (ii) the polling rate of RAPL energy reads.
The polling process transfers RAPL values from the kernel to user space through a call chain that can incur overhead at high sampling rates. Yet, such frequencies are necessary for fine-grained profiling at the function level.
We hypothesize that simplifying the tool design and shortening the call chain can reduce measurement overhead. To test this, we conduct two experiments: (i) a comparison of five existing tools and two lightweight implementations across six NAS benchmark functions, and (ii) a microbenchmark measuring the cost of key instructions in the call chain.
Results show measurable overhead in some tools. CodeCarbon, for example, adds up to 46.75\%, depending on the function profile. Simpler tools achieve high-frequency profiling with negligible impact. We also find that system calls are less efficient than the \texttt{rdmsr} instruction for accessing RAPL counters, introducing extra delay.
In the future, we plan to repeat the experiment including more benchmarks, including more complex software systems, and compare our solution to those based on eBPF \cite{geerd2025powerletrics}, a technology to execute custom programs within the operating system kernel to monitor system performance without requiring kernel modifications.
We aim to make \refuser and \refkernel a fully fledged tool to measure software energy usage at high frequency and therefore fully experiment on it, with different cache sizes.

\bibliographystyle{ACM-Reference-Format}
\bibliography{references}
\end{document}